\documentclass[aps,prl,twocolumn,showpacs,preprintnumbers,amsmath,amssymb,floatfix]{revtex4-1}
\usepackage{graphicx}

\begin{document}
\title{Lines of Fisher's zeros as separatrices for complex renormalization group flows}
\author{Yuzhi Liu}
\author{Y. Meurice}
\affiliation{Department of Physics and Astronomy, The University of Iowa, 
Iowa City, Iowa 52242, USA }
\date{\today}

\begin{abstract}
We extend the renormalization group transformation based on the two-lattice matching  to the  complex inverse temperature plane for Dyson's hierarchical Ising model. We consider values of the dimensional parameter above, below and exactly at the critical value where 
the ordered low temperature phase becomes impossible for a real positive temperature. We show numerically that, as the volume increases, the Fisher's  zeros appear to accumulate along lines that separate the flows ending on different fixed points.  We justify these findings in terms of  finite size scaling.  We argue that the location of the Fisher's zeros  at large volume determine the phase diagram in the complex plane.  We discuss the implications for nontrivial infrared fixed points in lattice gauge theory. 
\end{abstract}
\maketitle
\def\nm{{n_{max}}}
Classical field equations suggest the existence of massless Nambu-Goldstone modes in low temperature spin models with continuous symmetries, or massless gauge bosons 
in weakly coupled models with local symmetries. However, in some cases, quasiparticle excitations or instantons destroy the long range-order and generate a mass gap \cite{Polyakov1977429}.
This theoretical framework provides a justification for the absence of long-range order for the $D=1$ Ising model or the $D=2$ nonlinear $O(3)$ sigma model with nearest neighbor interactions, and the $D=3$  $U(1)$ or  $D=4$ $SU(2)$ lattice gauge theories with a compact Wilson's action.

For gauge theories, the absence of long-range order is associated with confinement. 
The possibility of almost losing confinement by adding a suitable number of fermions, in order to have a ``walking" coupling constant  has been considered for models of electro-weak symmetry breaking beyond the standard 
model \cite{Sannino:2009za}. This has motivated an intense activity in the lattice gauge theory community  \cite{Ogilvie:2010vx,DeGrand:2010ba}. From this point of view, it is crucial to decide unambiguously if an unusual infrared behavior will be present for a given number of fermion fields.  More generally, proving the (non)existence of a mass gap is often important and difficult  for condensed matter and particle physics models. 

\begin{figure}[h]
\includegraphics[width=3.3in,height=5.in,angle=0]{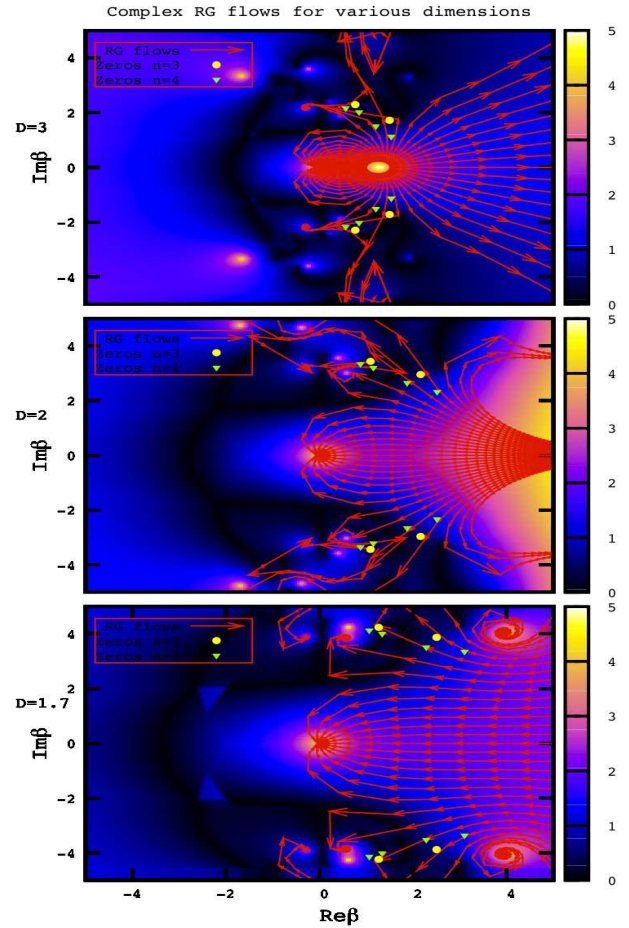}
\caption{RG flows in the complex $\beta$ plane  for $D=3$ (top), 2 (middle) and 1.7 (bottom) and Fisher's zeros for $\nm$ = 3 and 4. A darker  background indicates an ambiguous solution.}
\label{fig:HMflow}
\end{figure}

In the renormalization group (RG) approach, confinement means that the RG flows go uninterrupted from the weak coupling  region to the 
strong coupling region \cite{Ogilvie:2010vx,Tomboulis:2009zz}.
Recently, the loss of long-range order or the appearance of confinement when a parameter is varied has been explained in terms of RG fixed points disappearing 
in the complex  coupling or temperature plane \cite{Kaplan:2009kr,moroz09,Letter10,PhysRevD.83.056009}.
In this Letter, we provide a direct illustration of this mechanism  with numerical calculations of the RG flows in the complex inverse temperature ($\beta$)  plane using the two-lattice matching procedure \cite{PhysRevB.27.1736,Hasenfratz:1984hx} for Dyson's  hierarchical model \cite{dyson69,baker72} with an Ising measure. 
The choice of this model is justified by the feasibility of difficult numerical calculations explained below. 
The complex RG flows for three values of $D$ are illustrated  in Fig. \ref{fig:HMflow}.

From a practical point of view, our main point is that 
as the volume increases, the zeros of the partition function in the $\beta$ plane (Fisher's  zeros) accumulate along lines that separate the flows ending on different fixed points and thus determine the complex phase diagram. 

The Hamiltonian of  Dyson's hierarchical model with $2^\nm$ sites is: 
\begin{equation}
H =
-{\frac{1}{2}}\sum_{n=1}^{\nm}({\frac{c}{4}})^n  \sum_{B^{(n)}}
(\sum  _{x \in  B^{(n)} } \phi_x)^{2} \ 
\label{eq:hamiltonian}
\end{equation}
where $B^{(n)}$ denotes hierarchically nested blocks of size $2^n$. For details, we refer to a recent review article \cite{hmreview}. 
The parameter $c$ controls the decay of the interactions with the size of the blocks. From the scaling of a free massless Gaussian field under a change of the lattice spacing by a scaling factor $b$, we can include a dimension $D$ through the relation $c/4=b^{-2-D}$. Since the number of sites is divided by 2 at every blockspin transformation, we have  $b^D=2$.  
For $D=3$ ($c=2^{1/3}$), and more generally for $D>2$, the Ising hierarchical model has a second order phase transition and has many common features with the $D=3$ nearest neighbor interaction Ising model, however the critical exponent $\eta$ is zero. 
For $D=2$ ($c=1$), and more generally for $D\leq 2$ the model has no phase transition at finite temperature \cite{dyson69} unlike its $D=2$  nearest neighbor interaction counterpart. Our numerical study will focus on  $D=3$, 2 and 1.7.  The case $D=2$ is at the boundary and is quite interesting, as are other models playing this role (for instance, the $D=2$ nonlinear $O(2)$ sigma model with nearest neighbor interactions).

The partition function is obtained by integrating ${e^{-\beta H}}$ with a local measure $W_0(\phi)$. It can be calculated iteratively because the block variables are not mixed in $H$. If we call 
$W_n(\phi)$  the unnormalized, unrescaled, probability distribution of $\phi$ in a block of size $2^n$, we obtain the (implicitly $\beta$-dependent) recursion relation :
  \begin{equation}
\label{eq:iter}	
  W_{n+1}(\phi )=
   \int d\xi W_n(\frac{\phi}{2}+\xi)W_n(\frac{\phi}{2}-\xi) e^{\frac{\beta}{2}(\frac{c}{4})^{n+1}\phi^2}\ .
   \end{equation}

In the following, we consider an Ising measure $W_0(\phi) = \delta(\phi^2-1)$ and Eq. (\ref{eq:iter}) reduces to finite sums. This 
allows an exact calculation of  $W_n(\phi)$ as a {\it function} of $\beta$ as long as we have enough computer memory to keep track of the terms. 
At finite volume, the partition function is an entire function of the form: 
\begin{equation}
Z[\beta]=\sum_kN_k e^{{E_k}\beta}\ ,
\label{eq:partition}
\end{equation}
The zeros of $Z$ can be obtained from the intersections of the zero level curves for the real and imaginary parts
or from logarithmic residue methods used in Ref. \cite{Meurice:2009bq}.  

\begin{figure}[htp]
\includegraphics[width=3.3in,angle=0]{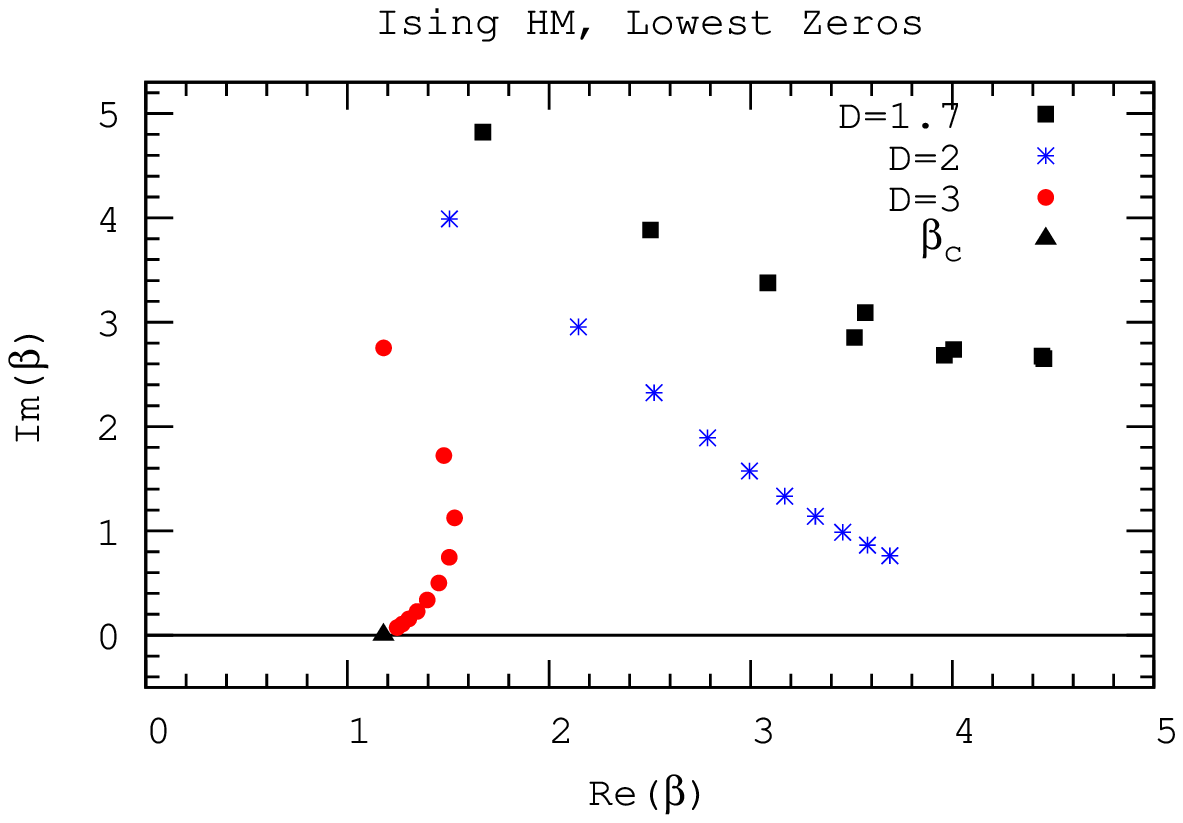}
\includegraphics[width=3.3in,angle=0]{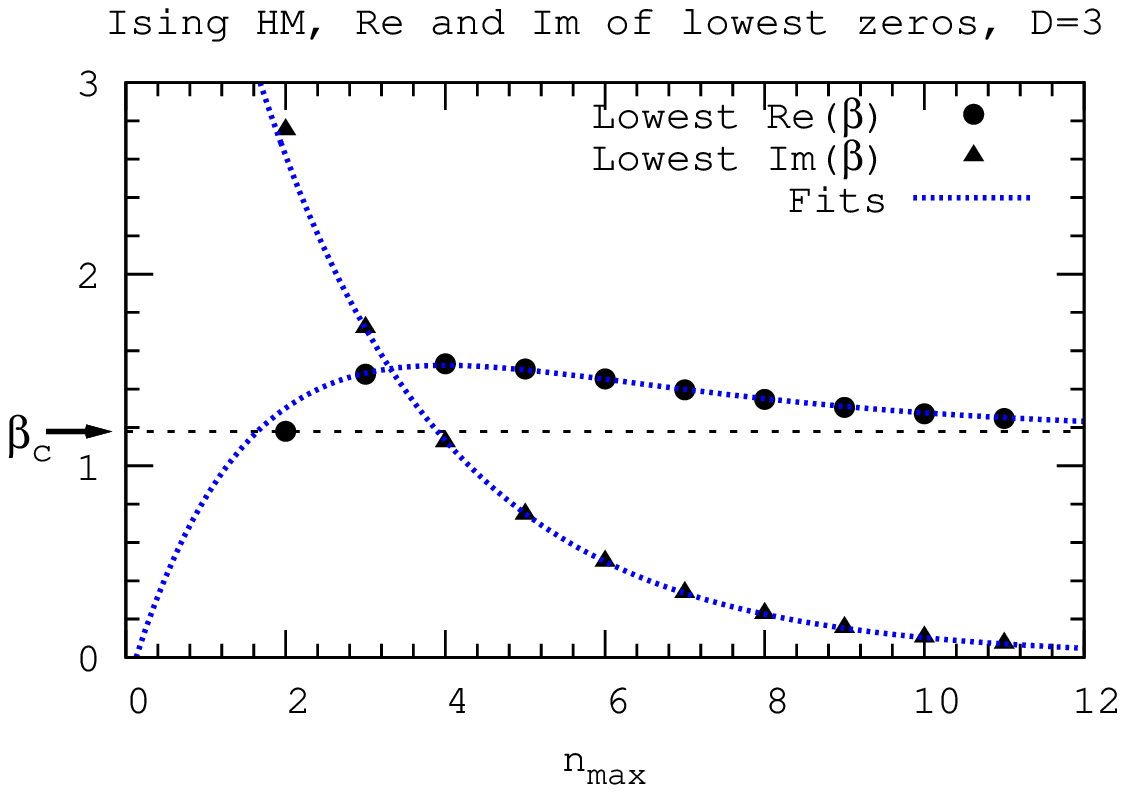}
\caption{The  lowest Fisher's zeros for $D=3$ (filled circle ending on $\beta _c$), 2 (crosses) and 1.7 (filled square) with $\nm$ from 2 to 11. As $\nm$ increases, the imaginary part decreases for $D$ =3 and 2 (top panel). Real and imaginary parts of Fisher's zeros for $D=3$ and $\nm$ going from 2 to 11 (bottom panel). }
\label{fig:hmzerod2}
\end{figure}

In Fig. \ref{fig:hmzerod2}, we compare  the lowest (closest to the real axis) Fisher's zeros for different volumes. The distribution of the Fisher's zeros is symmetric about the real axis and we only show the zeros in the upper half plane. 
For the $D=3$ case, the system has a second order phase transition at $\beta_c\simeq 1.179$ \cite{hmreview} and we can check the consistency of our results with finite size scaling \cite{Itzykson83,janke00}.  In Fig. \ref{fig:hmzerod2}, the lowest Fisher's zeros accumulate toward $\beta_c$ as the volume increases. The departure from a linear behavior is significant and requires  subleading corrections.  The singular part of the free energy can only depend on the linear size of the system $L$ through the combination $K_i L^{y_i}$, where $K_i$  are the nonlinear scaling variables. Under a RG transformation, $L\rightarrow L/b$ and $K_i\rightarrow b^{y_i}K_i$. If we only keep the relevant variable ($y_1=1/\nu)$ and the first irrelevant variable ($y_2=-\omega$), the requirement that $\beta_1(L)$ is the (lowest) zero imposes a relation of the form 
\begin{equation}
	KL^{1/\nu}= A+BL^{-\omega}+{\mathcal O}(L^{-2\omega})
\label{eq:scaling}
\end{equation}
Using the approximation $K\simeq  \beta_1(L)-\beta_c$ for the lowest zero, we obtain
\begin{eqnarray}
\label{eq:imscaling}
Re(\beta_1(L))-\beta_c &=& Re(A) L^{-1/\nu} + Re(B) L^{-1/\nu-\omega} \label{eq:rescaling}\\ \nonumber
	Im(\beta_1(L))&=& Im(A) L^{-1/\nu} + Im(B) L^{-1/\nu-\omega}
	\end{eqnarray}
Using the known values for $\beta_c$, $\nu$, and $\omega$ \cite{hmreview} and fitting for $A$ and $B$, we obtain results displayed in Fig. \ref{fig:hmzerod2}. 
The fits are in very good agreement with the data except for the smallest volume. 

Complex RG flows have been studied and discussed for a variety of statistical mechanics, many-body and lattice models
\cite{Damgaard:1993df,PhysRevE.52.4512,moroz09,Letter10,PhysRevD.83.056009}. 
Here we construct the RG flows in the complex $\beta$ plane by using a two-lattice matching method inspired by Refs. \cite{PhysRevB.27.1736,Hasenfratz:1984hx}. 
At level $n$ in the block hierarchy, we split the system into two blocks $B_1$ and $B_2$ each with $2^{n-1}$ sites. The observable we considered is 
 \begin{equation}
R(\beta,n)\equiv\frac{\left\langle (\sum_{x\in B_1}\phi_x )(\sum_{y\in B_2}\phi_y)\right\rangle_{\beta , n }}{\left\langle (\sum_{x\in B_1}\phi_x)(\sum_{y\in B_1}\phi_y ))\right\rangle_{\beta , n}} 
\label{eq:obs}
\end{equation}
This ratio measures how the blocks are correlated. It does not require any division by the partition function. The field rescaling which needs to be calculated to write down a full RG  transformation cancels  out. 
Using $W_{n-1}(\phi)$ for the two block variables, we obtain
\begin{eqnarray}
&\ &R(\beta,n)= \\ \nonumber
&\ &  \frac{\int d\phi_1 d\phi_2 e^{\frac{\beta}{2}(\frac{c}{4})^{n}(\phi_1+\phi_2)^2}\phi_1\phi_2 W_{n-1}(\phi_1)W_{n-1}(\phi_2)}{\int d\phi_1 d\phi_2 e^{\frac{\beta}{2}(\frac{c}{4})^{n}(\phi_1+\phi_2)^2}{\phi_1}^2 W_{n-1}(\phi_1)W_{n-1}(\phi_2)}
\end{eqnarray}
Unlike RG transformations based on decimation for one-dimensional Ising models \cite{Damgaard:1993df,PhysRevE.52.4512},  
the measures obtained by iterating Eq. (\ref{eq:iter}) are not Ising measures. Instead, we rely on the 
assumption that after enough RG transformations, the flows become {\it approximately} one-dimensional and that we can use $\beta$ as a coordinate for the unstable direction. 
The matching condition reads: 
\begin{equation}
R(\beta,n)=R(\beta',n-1)\  ,
\label{eq:matching}
\end{equation}
with $n$ as large as possible. 
Given an initial $\beta$, the numerical calculation of $\beta'$ can be done with the methods used to find Fisher's zeros.  

\begin{figure}
\includegraphics[width=3.3in,angle=0]{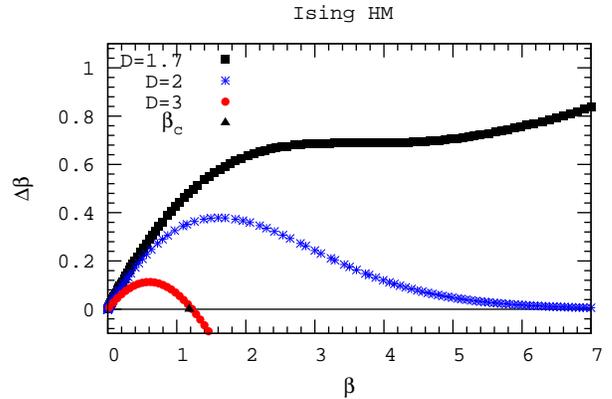}
\caption{Discrete $\beta$ function for $D$= 3, 2 and 1.7\label{fig:db}}
\end{figure}

This procedure yields very simple flows for real values of $\beta$. They can be described in terms of 
a discrete Callan-Symanzik $\beta$ function defined as: 
\begin{equation}
\Delta\beta(\beta)= \beta - \beta^\prime \  .
\end{equation}
\def\db{\Delta\beta}
The existence of zeros of $\Delta\beta$  implies the existence of fixed points. Figure \ref{fig:db} shows $\db$ for $D=3,\ 2$ and 1.7.  It is not surprising that there is only one zero at the origin for $D=$ 2 and 1.7 since it has been proven \cite{dyson69} that there is no phase transition at finite temperature. For $D=3$, another nontrivial zero appears near $\beta _c$. As the volume increases, the nontrivial zeros approach $\beta_c$. 

For $D=2$,  a low temperature expansion can be used to show that at leading order
\begin{equation}
\Delta\beta({\beta})\propto e^{-B_n\beta} \  . 
\end{equation}
 The lowest order contributions to $R(\beta,n)$ for $D=2$ comes from flipping all the spins in one of the blocks of size $2^{n-1}$. The energy cost is only -1/2 and is $n$-independent. The leading order contribution to $\db$ comes from the first $n$-dependent energy cost. Numerically, 
$B_3=35/32$, $B_4=155/128$ and $B_5=651/512$. The general formula is $B_n=((11+1/4^{n-1})/6) -1/2^{n-1}-1/2$ and $B_\infty = 4/3$. The exponential decay  is in contrast with 
$\Delta\beta$ for $D=2$ nearest neighbor $O(N)$ models in the large $N$ limit which has been calculated for the two-lattice matching \cite{PhysRevB.27.1736}  and a simple rescaling of cutoff in the gap equation \cite{PhysRevD.83.056009},  where at infinite volume, $\db$ approaches ${\ln b}/{(2\pi)}$ asymptotically. 
It is possible to find simple models for the continuous Callan-Symanzik $\beta$-function with continuous parameters 
that interpolate qualitatively among the various behaviors of $\db$ in Fig. \ref{fig:db} and that have zeros moving in the complex plane when 
$D=2$. Functional conjugation methods \cite{PhysRevD.83.065019} provide a more systematic approach of the relationship. 

In the complex case, it appears in general the matching equation has more than one solution. 
For $D=2$, there is a fixed number of solutions, because the matching equation reduces to a polynomial equation (the $E_k$ in Eq. (\ref{eq:partition}) are all rational numbers).
In order to resolve this ambiguity, we picked the $\beta'$ that minimizes $|\beta-\beta'|$. Under some circumstances, there is only one $\beta'$ close to $\beta$ and this seems quite natural but this is not always the case. In order to quantify this ambiguity, we defined:
\begin{equation}
f(\beta)\equiv \log |\beta-\beta'_2|-\log |\beta-\beta'_1|\   ,
\label{eq:ambi}
\end{equation}
where $\beta'_1$ is the closest solution and $\beta_2'$ the second closest. 
In Fig. \ref{fig:HMflow}, we made a contour plot of $f(\beta)$.  The darker the color, the more ambiguous is the selection of $\beta'$. In dark regions, the flow is rather erratic. This reflects the existence of several competing solutions rather than some intrinsic ``chaos" as for the $D=1$ Ising model \cite{PhysRevE.52.4512}. On the other hand, the mostly unambiguous flows of 
Fig. \ref{fig:HMflow} show simple patterns. 
For $D=3$, the zeros appear near the boundary of the basins of attraction the two stable fixed points at 0 and $\infty$. As $D$ decreases, the fixed point moves to larger $\beta$ and becomes infinite at $D=2$. 
For lower $D$, a pair of complex conjugated zeros appears. 

\begin{figure}
\includegraphics[width=3.3in,angle=0]{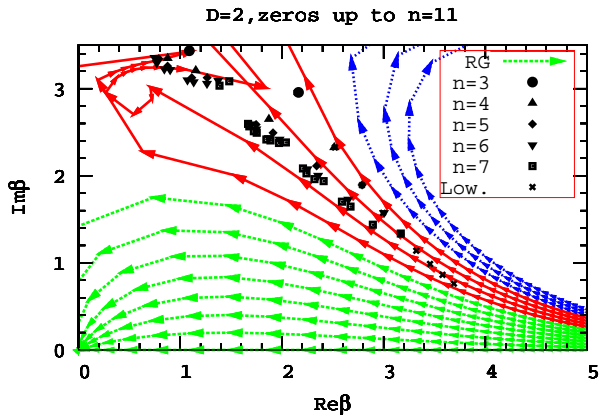}
\caption{\label{fig:small} Detail of Fig. \ref{fig:HMflow} for $D=2$ and Fisher's zeros for larger number of sites described in the text. }
\end{figure}

Figure \ref{fig:HMflow} indicates that Fisher's zeros for the two volumes matched appear in regions where the flows are ambiguous and at the end of flow lines that neither go to zero or other fixed points. The zeros for larger volumes move approximately ``backward" along the set of 
flow lines that separate the flows going to different fixed points. This is illustrated  in Fig. \ref{fig:small} for $D=2$, where all the zeros corresponding to volumes up to $2^7$ fitting in the frame and the lowest zeros for volumes up to $2^{11}$ have been displayed. 
This behavior is in approximate agreement  with the argument \cite{Damgaard:1993df}  that under a RG transformation, Fisher's zeros for $2^n$ sites should map into Fisher's zeros for $2^{n-1}$ sites.  In view of this argument, the general shape 
of the flows and the absence of transition on the real axis, it is plausible that as the volume increases, the line extends to infinity.  Lines of Fisher's zeros are quite common \cite{janke00}. Surfaces were also observed in the complex variable sinh($2\beta$) \cite{janke04}.

In summary, we have calculated the complex RG flows for three  values of $D$. 
For $D=3$, the Fisher's zeros pinch the real axis at $\beta_c$ and separate the flows going to 0 (symmetric phase) and $\infty$ (broken symmetry phase). For $D\leq 2$, it is plausible that 
the line of Fisher's zero goes to infinity with no contact with the real axis. This hypothetically infinite line separates the flows going to 0 from those curling back to either a complex fixed point  or infinity. This suggests that the basins of attraction of the various fixed points in the complex plane are separated by lines of Fisher's zeros. From a practical point of view, it is easier to calculate Fisher's zeros than to construct complex RG flows. A direct interpretation of complex flow would be desirable.
In problems involving oscillations, the addition of an imaginary part to the energy or frequency corresponds to dissipation, skin depth or a finite 
lifetime. A simple example is the electric permittivity $\epsilon$ becoming complex as a result of  damping. From this point of view, 
a complex energy density $(\epsilon/2) E^2$ is an effective description of the interactions with matter. Complex eigenvalues for the transfer matrix were also found 
\cite{Meisinger:2010be} and could suggest a more complete picture of complex RG flows. 

We thank the participants of the workshop 
``New applications of the RG 
method" held at the Institute for Nuclear Theory, University of Washington, Seattle and of the workshop ``Critical Behavior of Lattice models"
at the Aspen Center for Physics in May and June 2010 for stimulating discussions. This 
research was supported in part  by the Department of Energy
under Contract No. FG02-91ER40664.


\begin{thebibliography}{10}%
\makeatletter
\providecommand \@ifxundefined [1]{%
 \ifx #1\undefined \expandafter \@firstoftwo
 \else \expandafter \@secondoftwo
\fi
}%
\providecommand \@ifnum [1]{%
 \ifnum #1\expandafter \@firstoftwo
 \else \expandafter \@secondoftwo
\fi
}%
\providecommand \enquote [1]{``#1''}%
\providecommand \bibnamefont  [1]{#1}%
\providecommand \bibfnamefont [1]{#1}%
\providecommand \citenamefont [1]{#1}%
\providecommand\href[0]{\@sanitize\@href}%
\providecommand\@href[1]{\endgroup\@@startlink{#1}\endgroup\@@href}%
\providecommand\@@href[1]{#1\@@endlink}%
\providecommand \@sanitize [0]{\begingroup\catcode`\&12\catcode`\#12\relax}%
\@ifxundefined \pdfoutput {\@firstoftwo}{%
 \@ifnum{\z@=\pdfoutput}{\@firstoftwo}{\@secondoftwo}%
}{%
 \providecommand\@@startlink[1]{\leavevmode}%
 \providecommand\@@endlink[0]{}%
}{%
 \providecommand\@@startlink[1]{%
  \leavevmode
  \pdfstartlink
   attr{/Border[0 0 1 ]/H/I/C[0 1 1]}%
   user{/Subtype/Link/A<</Type/Action/S/URI/URI(#1)>>}%
  \relax
 }%
 \providecommand\@@endlink[0]{\pdfendlink}%
}%
\providecommand \url  [0]{\begingroup\@sanitize \@url }%
\providecommand \@url [1]{\endgroup\@href {#1}{\urlprefix}}%
\providecommand \urlprefix [0]{URL }%
\providecommand \Eprint[0]{\href }%
\@ifxundefined \urlstyle {%
  \providecommand \doi [1]{doi:\discretionary{}{}{}#1}%
}{%
  \providecommand \doi [0]{doi:\discretionary{}{}{}\begingroup
  \urlstyle{rm}\Url }%
}%
\providecommand \doibase [0]{http://dx.doi.org/}%
\providecommand \Doi[1]{\href{\doibase#1}}%
\providecommand \bibAnnote [3]{%
  \BibitemShut{#1}%
  \begin{quotation}\noindent
    \textsc{Key:}\ #2\\\textsc{Annotation:}\ #3%
  \end{quotation}%
}%
\providecommand \bibAnnoteFile [2]{%
  \IfFileExists{#2}{\bibAnnote {#1} {#2} {\input{#2}}}{}%
}%
\providecommand \typeout [0]{\immediate \write \m@ne }%
\providecommand \selectlanguage [0]{\@gobble}%
\providecommand \bibinfo [0]{\@secondoftwo}%
\providecommand \bibfield [0]{\@secondoftwo}%
\providecommand \translation [1]{[#1]}%
\providecommand \BibitemOpen[0]{}%
\providecommand \bibitemStop [0]{}%
\providecommand \bibitemNoStop [0]{.\EOS\space}%
\providecommand \EOS [0]{\spacefactor3000\relax}%
\providecommand \BibitemShut [1]{\csname bibitem#1\endcsname}%
\bibitem{Polyakov1977429}%
  \BibitemOpen
  \bibfield{author}{%
  \bibinfo {author} {\bibfnamefont{A.~M.}\ \bibnamefont{Polyakov}},\ }%
  \bibfield{journal}{%
  \Doi{DOI: 10.1016/0550-3213(77)90086-4}{\bibinfo {journal} {Nucl.  Phys. 
  B}}\ }%
  \textbf{\bibinfo {volume} {120}},\ \bibinfo {pages} {429 } (\bibinfo {year}
  {1977})
  \bibAnnoteFile{NoStop}{Polyakov1977429}%
\bibitem{Sannino:2009za}%
  \BibitemOpen
  \bibfield{author}{%
  \bibinfo {author} {\bibfnamefont{F.}~\bibnamefont{Sannino}},\ }%
  \bibfield{journal}{%
  \bibinfo {journal} {Acta Phys. Polon.}\ }%
  \textbf{\bibinfo {volume} {B40}},\ \bibinfo {pages} {3533} (\bibinfo {year}
  {2009})
  \bibAnnoteFile{NoStop}{Sannino:2009za}%
 \bibitem{Ogilvie:2010vx}%
  \BibitemOpen
  \bibfield{author}{%
  \bibinfo {author} {\bibfnamefont{M.~C.}\ \bibnamefont{Ogilvie}},\ }%
  \bibfield{journal}{%
  \bibinfo {journal} {Phil. Trans. Roy. Soc. Lond.}\ }%
  \textbf{\bibinfo {volume} {A }},\ \bibinfo {pages} {in press} (\bibinfo {year}
  {2011}),\ \Eprint{http://arxiv.org/abs/1010.1942}{arXiv:1010.1942 [hep-lat]}%
  \bibAnnoteFile{NoStop}{Ogilvie:2010vx}%
\bibitem{DeGrand:2010ba}%
  \BibitemOpen
  \bibfield{author}{%
  \bibinfo {author} {\bibfnamefont{T.}~\bibnamefont{DeGrand}},\ }%
  \bibfield{journal}{%
  \bibinfo {journal} {Phil. Trans. Roy. Soc. Lond.}\ }%
  \textbf{\bibinfo {volume} {A }},\ \bibinfo {pages} {in press} (\bibinfo {year}
  {2011}),\ \Eprint{http://arxiv.org/abs/1010.4741}{arXiv:1010.4741 [hep-lat]}%
  \bibAnnoteFile{NoStop}{DeGrand:2010ba}%
\bibitem{Tomboulis:2009zz}%
  \BibitemOpen
  \bibfield{author}{%
  \bibinfo {author} {\bibfnamefont{E.~T.}\ \bibnamefont{Tomboulis}},\ }%
  \bibfield{journal}{%
  \Doi{10.1142/S0217732309032307}{\bibinfo {journal} {Mod. Phys. Lett.}}\ }%
  \textbf{\bibinfo {volume} {A24}},\ \bibinfo {pages} {2717} (\bibinfo {year}
  {2009})%
  \bibAnnoteFile{NoStop}{Tomboulis:2009zz}%
\bibitem{Kaplan:2009kr}%
  \BibitemOpen
  \bibfield{author}{%
  \bibinfo {author} {\bibfnamefont{D.~B.}\ \bibnamefont{Kaplan}}, \bibinfo
  {author} {\bibfnamefont{J.-W.}\ \bibnamefont{Lee}}, \bibinfo {author}
  {\bibfnamefont{D.~T.}\ \bibnamefont{Son}},\ and\ \bibinfo {author}
  {\bibfnamefont{M.~A.}\ \bibnamefont{Stephanov}},\ }%
  \bibfield{journal}{%
  \Doi{10.1103/PhysRevD.80.125005}{\bibinfo {journal} {Phys. Rev.}}\ }%
  \textbf{\bibinfo {volume} {D80}},\ \bibinfo {pages} {125005} (\bibinfo {year}
  {2009})
  \bibAnnoteFile{NoStop}{Kaplan:2009kr}%
\bibitem{moroz09}%
  \BibitemOpen
  \bibfield{author}{%
  \bibinfo {author} {\bibfnamefont{S.}~\bibnamefont{Moroz}}\ and\ \bibinfo
  {author} {\bibfnamefont{R.}~\bibnamefont{Schmidt}},\ }%
  \bibfield{journal}{%
  \Doi{10.1016/j.aop.2009.10.002}{\bibinfo {journal} {Annals Phys.}}\ }%
  \textbf{\bibinfo {volume} {325}},\ \bibinfo {pages} {491} (\bibinfo {year}
  {2010})
  \bibAnnoteFile{NoStop}{moroz09}%
\bibitem{Letter10}%
  \BibitemOpen
  \bibfield{author}{%
  \bibinfo {author} {\bibfnamefont{A.}~\bibnamefont{{Denbleyker}}}, \bibinfo
  {author} {\bibfnamefont{D.}~\bibnamefont{{Du}}}, \bibinfo {author}
  {\bibfnamefont{Y.}~\bibnamefont{{Liu}}}, \bibinfo {author}
  {\bibfnamefont{Y.}~\bibnamefont{{Meurice}}},\ and\ \bibinfo {author}
  {\bibfnamefont{H.}~\bibnamefont{{Zou}}},\ }%
  \bibfield{journal}{%
  \Doi{10.1103/PhysRevLett.104.251601}{\bibinfo {journal} {Phys. Rev. 
  Lett.}}\ }%
  \textbf{\bibinfo {volume} {104}},\ \bibinfo {pages} {251601} (
  \bibinfo {year} {2010})
  \bibAnnoteFile{NoStop}{Letter10}%
\bibitem{PhysRevD.83.056009}%
  \BibitemOpen
  \bibfield{author}{%
  \bibinfo {author} {\bibfnamefont{Y.}~\bibnamefont{Meurice}}\ and\ \bibinfo
  {author} {\bibfnamefont{H.}~\bibnamefont{Zou}},\ }%
  \bibfield{journal}{%
  \Doi{10.1103/PhysRevD.83.056009}{\bibinfo {journal} {Phys. Rev. D}}\ }%
  \textbf{\bibinfo {volume} {83}},\ \bibinfo {pages} {056009} (
  \bibinfo {year} {2011})%
  \bibAnnoteFile{NoStop}{PhysRevD.83.056009}%
\bibitem{PhysRevB.27.1736}%
  \BibitemOpen
  \bibfield{author}{%
  \bibinfo {author} {\bibfnamefont{J.~E.}\ \bibnamefont{Hirsch}}\ and\ \bibinfo
  {author} {\bibfnamefont{S.~H.}\ \bibnamefont{Shenker}},\ }%
  \bibfield{journal}{%
  \Doi{10.1103/PhysRevB.27.1736}{\bibinfo {journal} {Phys. Rev. B}}\ }%
  \textbf{\bibinfo {volume} {27}},\ \bibinfo {pages} {1736} (
  \bibinfo {year} {1983})%
  \bibAnnoteFile{NoStop}{PhysRevB.27.1736}%
\bibitem{Hasenfratz:1984hx}%
  \BibitemOpen
  \bibfield{author}{%
  \bibinfo {author} {\bibfnamefont{A.}~\bibnamefont{Hasenfratz}}, \bibinfo
  {author} {\bibfnamefont{P.}~\bibnamefont{Hasenfratz}}, \bibinfo {author}
  {\bibfnamefont{U.~M.}\ \bibnamefont{Heller}},\ and\ \bibinfo {author}
  {\bibfnamefont{F.}~\bibnamefont{Karsch}},\ }%
  \bibfield{journal}{%
  \Doi{10.1016/0370-2693(84)91051-7}{\bibinfo {journal} {Phys. Lett.}}\ }%
  \textbf{\bibinfo {volume} {B140}},\ \bibinfo {pages} {76} (\bibinfo {year}
  {1984})%
  \bibAnnoteFile{NoStop}{Hasenfratz:1984hx}%
\bibitem{dyson69}%
  \BibitemOpen
  \bibfield{author}{%
  \bibinfo {author} {\bibfnamefont{F.}~\bibnamefont{Dyson}},\ }%
  \bibfield{journal}{%
  \bibinfo {journal} {Comm.\ Math.\ Phys.}\ }%
  \textbf{\bibinfo {volume} {12}},\ \bibinfo {pages} {91} (\bibinfo {year}
  {1969})%
  \bibAnnoteFile{NoStop}{dyson69}%
\bibitem{baker72}%
  \BibitemOpen
  \bibfield{author}{%
  \bibinfo {author} {\bibfnamefont{G.}~\bibnamefont{Baker}},\ }%
  \bibfield{journal}{%
  \bibinfo {journal} {Phys.\ Rev.\ B}\ }%
  \textbf{\bibinfo {volume} {5}},\ \bibinfo {pages} {2622} (\bibinfo {year}
  {1972})%
  \bibAnnoteFile{NoStop}{baker72}%
\bibitem{hmreview}%
  \BibitemOpen
  \bibfield{author}{%
  \bibinfo {author} {\bibfnamefont{Y.}~\bibnamefont{Meurice}},\ }%
  \bibfield{journal}{%
  \bibinfo {journal} {J. Phys.}\ }%
  \textbf{\bibinfo {volume} {A40}},\ \bibinfo {pages} {R39} (\bibinfo {year}
  {2007})
  \bibAnnoteFile{NoStop}{hmreview}%
\bibitem{Meurice:2009bq}%
  \BibitemOpen
  \bibfield{author}{%
  \bibinfo {author} {\bibfnamefont{Y.}~\bibnamefont{Meurice}},\ }%
  \bibfield{journal}{%
  \bibinfo {journal} {Phys. Rev.}\ }%
  \textbf{\bibinfo {volume} {D80}},\ \bibinfo {pages} {054020} (\bibinfo {year}
  {2009}) 
  \bibAnnoteFile{NoStop}{Meurice:2009bq}%
\bibitem{Itzykson83}%
  \BibitemOpen
  \bibfield{author}{%
  \bibinfo {author} {\bibfnamefont{C.}~\bibnamefont{{Itzykson}}}, \bibinfo
  {author} {\bibfnamefont{R.~B.}\ \bibnamefont{{Pearson}}},\ and\ \bibinfo
  {author} {\bibfnamefont{J.~B.}\ \bibnamefont{{Zuber}}},\ }%
  \bibfield{journal}{%
  \Doi{10.1016/0550-3213(83)90499-6}{\bibinfo {journal} {Nucl. Phys. B}}\
  }%
  \textbf{\bibinfo {volume} {220}},\ \bibinfo {pages} {415} (
  \bibinfo {year} {1983})%
  \bibAnnoteFile{NoStop}{Itzykson83}%
\bibitem{janke00}%
  \BibitemOpen
  \bibfield{author}{%
  \bibinfo {author} {\bibfnamefont{W.}~\bibnamefont{Janke}}\ and\ \bibinfo
  {author} {\bibfnamefont{R.}~\bibnamefont{Kenna}},\ }%
  \bibfield{journal}{%
  \bibinfo {journal} {J. Stat. Phys.}\ }%
  \textbf{\bibinfo {volume} {102}},\ \bibinfo {pages} {1211} (\bibinfo {year}
  {2001}) 
  \bibAnnoteFile{NoStop}{janke00}%
\bibitem{Damgaard:1993df}%
  \BibitemOpen
  \bibfield{author}{%
  \bibinfo {author} {\bibfnamefont{P.~H.}\ \bibnamefont{Damgaard}}\ and\
  \bibinfo {author} {\bibfnamefont{U.~M.}\ \bibnamefont{Heller}},\ }%
  \bibfield{journal}{%
  \Doi{10.1016/0550-3213(93)90526-U}{\bibinfo {journal} {Nucl. Phys.}}\ }%
  \textbf{\bibinfo {volume} {B410}},\ \bibinfo {pages} {494} (\bibinfo {year}
  {1993})
  \bibAnnoteFile{NoStop}{Damgaard:1993df}%
\bibitem{PhysRevE.52.4512}%
  \BibitemOpen
  \bibfield{author}{%
  \bibinfo {author} {\bibfnamefont{B.~P.}\ \bibnamefont{Dolan}},\ }%
  \bibfield{journal}{%
  \Doi{10.1103/PhysRevE.52.4512}{\bibinfo {journal} {Phys. Rev. E}}\ }%
  \textbf{\bibinfo {volume} {52}},\ \bibinfo {pages} {4512} (
  \bibinfo {year} {1995})%
  \bibAnnoteFile{NoStop}{PhysRevE.52.4512}%
\bibitem{PhysRevD.83.065019}%
  \BibitemOpen
  \bibfield{author}{%
  \bibinfo {author} {\bibfnamefont{T.~L.}\ \bibnamefont{Curtright}}\ and\
  \bibinfo {author} {\bibfnamefont{C.~K.}\ \bibnamefont{Zachos}},\ }%
  \bibfield{journal}{%
  \Doi{10.1103/PhysRevD.83.065019}{\bibinfo {journal} {Phys. Rev. D}}\ }%
  \textbf{\bibinfo {volume} {83}},\ \bibinfo {pages} {065019} (
  \bibinfo {year} {2011})%
  \bibAnnoteFile{NoStop}{PhysRevD.83.065019}%
\bibitem{janke04}%
  \BibitemOpen
  \bibfield{author}{%
  \bibinfo {author} {\bibfnamefont{W.}~\bibnamefont{{Janke}}}, \bibinfo
  {author} {\bibfnamefont{D.~A.}\ \bibnamefont{{Johnston}}},\ and\ \bibinfo
  {author} {\bibfnamefont{R.}~\bibnamefont{{Kenna}}},\ }%
  \bibfield{journal}{%
  \Doi{10.1016/j.nuclphysb.2004.01.028}{\bibinfo {journal} {Nucl. Phys. 
  B}}\ }%
  \textbf{\bibinfo {volume} {682}},\ \bibinfo {pages} {618} (
  \bibinfo {year} {2004})
  \bibAnnoteFile{NoStop}{janke04}%
  \bibitem{Meisinger:2010be}%
  \BibitemOpen
  \bibfield{author}{%
  \bibinfo {author} {\bibfnamefont{P.~N.}\ \bibnamefont{Meisinger}}, \bibinfo
  {author} {\bibfnamefont{M.~C.}\ \bibnamefont{Ogilvie}},\ and\ \bibinfo
  {author} {\bibfnamefont{T.~D.}\ \bibnamefont{Wiser}}}\ , %
  Int. J. Theor. Phys. 50: 1042Ð1051
   (\bibinfo {year} {2011})
  \bibAnnoteFile{NoStop}{Meisinger:2010be}%
\end{thebibliography}
%

\end{document}